\documentclass[aps,prd,reprint, longbibliography,notitlepage,nofootinbib,nobibnotes,superscriptaddress,amsmath,amssymb,preprintnumbers]{revtex4-2}

\usepackage{graphicx} 
\usepackage{dcolumn}
\usepackage{bm}        % for math
\usepackage{amssymb}   % for math
\usepackage{adjustbox}  
\usepackage{amsmath,array}
\usepackage{comment}
\usepackage{natbib}
\usepackage{MnSymbol}
\usepackage{caption}
\usepackage{subcaption}
\usepackage[colorlinks=true,linkcolor=blue,citecolor=blue, urlcolor=blue]{hyperref}
\usepackage[capitalise]{cleveref}
\usepackage[usenames,dvipsnames]{color}
\usepackage     [utf8]                  {inputenc}
\usepackage     [T1]                    {fontenc}
\usepackage     [english]               {babel}
\usepackage{epigraph}
\usepackage{etoolbox}
\usepackage{ dsfont }
\usepackage{physics}
\usepackage{float}
\usepackage{xcolor}
\usepackage{siunitx}
\makeatletter
\patchcmd{\epigraph}{\@epitext{#1}}{\itshape\@epitext{#1}}{}{}  
\newcommand*\eqsize{%
\@setfontsize\mysize{9.0}{9.0}%
    }
    \usepackage{multirow}
\makeatother

\usepackage{xfrac}
\usepackage{amsmath,amssymb}
\usepackage{esdiff} % derivatives
\usepackage{commath} % math macros
\usepackage{bbm} % blackboard style symbols
\usepackage{braket}
\usepackage{slashed}
\allowdisplaybreaks

\newcommand{\kompost}{K{\o}MP{\o}ST}
\newcommand{\wTilde}{\tilde{w}}

\newcommand{\pp}{\vec{p}}

\newcommand{\xT}{\mathbf{x}_T}			% transverse x, vector
\newcommand{\GeV}{\;\text{GeV}}

\newcommand{\hydro}{{\rm hydro}}

\newcommand{\p}{\mathbf{p}}
\newcommand{\x}{\mathbf{x}}

\newcommand{\q}{\mathbf{q}}

\newcommand{\dileptons}{l^{+}l^{-}}

\definecolor{oscar}{RGB}{22, 156, 172}

\definecolor{oscarC}{RGB}{22, 156, 172}

\makeatletter
\renewcommand\@makecaption[2]{%
  \par
  \vskip\abovecaptionskip
  \begingroup
   \small\rmfamily
    \begingroup
     \samepage
     \flushing
     \let\footnote\@footnotemark@gobble
     \@make@capt@title{#1}{#2}\par
    \endgroup
  \endgroup
  \vskip\belowcaptionskip
}
\makeatother

\begin{document}

\date{\today}

\title{Scaling of pre-equilibrium dilepton production in QCD Kinetic Theory}

\author{Oscar Garcia-Montero}
\email{garcia@physik.uni-bielefeld.de}
\affiliation{Fakult\"at f\"ur Physik, Universit\"at Bielefeld, D-33615 Bielefeld, Germany}

\author{Philip Plaschke}
\email{pplaschke@physik.uni-bielefeld.de}
\affiliation{Fakult\"at f\"ur Physik, Universit\"at Bielefeld, D-33615 Bielefeld, Germany}

\author{S\"oren Schlichting}
\email{sschlichting@physik.uni-bielefeld.de}
\affiliation{Fakult\"at f\"ur Physik, Universit\"at Bielefeld, D-33615 Bielefeld, Germany}

\begin{abstract} 
We use QCD kinetic theory to compute dilepton production coming from the pre-equilibrium phase of the Quark-Gluon Plasma created in high-energy heavy-ion collisions. We demonstrate that the dilepton spectrum exhibits a simple scaling in terms of the specific shear viscosity $\eta/s$ and entropy density $dS/d\zeta \sim {\scriptstyle \left(T\tau^{1/3}\right)_\infty^{3/2}}$, which can be derived from dimensional analysis in the presence of a pre-equilibrium attractor. Based on this scaling we perform event-by-event calculations of in-medium dilepton production and determine the invariant mass range where the pre-equilibrium yield is the leading contribution.
\end{abstract}

\maketitle

%~\cite{Mazeliauskas:2018yef,Berges:2020fwq,Du:2022bel,Du:2020dvp,Du:2020zqg}
\section{Introduction.}
Developing a first principles understanding of the non-equilibrium QCD processes that lead to the formation of a close to equilibrium Quark-Gluon Plasma (QGP), is one of the most outstanding challenges in modern heavy-ion physics. By now it has been established~\cite{Schlichting:2019abc,Berges:2020fwq} that -- in various different microscopic theories -- the onset of hydrodynamic behavior typically occurs on the time scale of a single equilibrium relaxation time $\sim(4\pi \eta/s)/T_{\rm eq}$, as determined by the specific shear viscosity $\eta/s$ and (local) temperature of the system. Over the course of this evolution, the non-equilibrium QGP also undergoes a rapid information loss, whereby (some) details of the initial conditions become irrelevant, such that the pre-equilibrium evolution is governed by pre-equilibrium attractors (see Ref.~\cite{Heller:2015dha,Florkowski:2017olj,Romatschke:2017ejr,Kurkela:2019set,Du:2020zqg,Soloviev:2021lhs} and references therein) even before a viscous fluid dynamic description becomes applicable.

Despite significant theoretical progress, experimental measurements to probe the early-time pre-equilibrium dynamics of the QGP at present and future facilities at RHIC and the LHC remain elusive. Electromagnetic probes, namely photons $(\gamma)$ and dileptons ($\dileptons$), provide promising candidates, as they are produced throughout the entire space-time evolution of the collisions, and due to the lack of re-interactions with the QGP medium, escape the reaction unscathed. However, experimental measurement that isolate this signal are extremely challenging, as in addition to the ``direct'' production, which can be further sub-divided into prompt contributions (Drell-Yan~\cite{Drell:1970wh}), thermal photons/dileptons and early-time pre-equilibrium radiation of the QGP, photons and dileptons are also copiously produced from late stage hadronic decays (see e.g.~\cite{Rapp:2013nxa,Vujanovic:2016anq} and references therein).

Evidently, an important property of dileptons is the invariant mass $M$ of the lepton pair, as it can be used to discriminate between different production mechanism~\cite{Coquet:2023wjk,Coquet:2021lca}. Generally speaking, lepton pairs with larger invariant masses tend to be produced at earlier times, such that e.g. the Drell-Yan process is the main source of high-$M$ dileptons. The intermediate mass region is produced by in-medium dileptons coming from the pre-equilibrium and thermal phase of the QGP. Recent estimates of the pre-equilibrium dilepton production have shown that in an invariant mass range of $3\GeV < M < 4\GeV$ the pre-equilibrium contribution can indeed exceed the thermal QGP and Drell-Yan contributions~\cite{Coquet:2021lca,Coquet:2021gms}, and first ideas to develop a phenomenology of the pre-equilibrium phase based on dilepton measurements have been developed~\cite{Coquet:2023wjk,Seck:2023oyt}. 

In this work we show that the pre-equilibrium dilepton production exhibits a simple scaling behavior (c.f. \cref{eq:SpectrumScaling}) in terms of the specific shear viscosity $\eta/s$ and local entropy density $dS/d^2\xT d\zeta$ of the QGP, which can also be extended to other probes of the early pre-equilibrium phase~\cite{Garcia-Montero:2023lrd}. By performing microscopic calculations in QCD kinetic theory (KT)~\cite{Arnold:2002ja,Arnold:2002zm,Arnold:2001ms,Arnold:2001ba}, which simultaneously provides a consistent theoretical description of the thermalization of the QGP and the production of electromagnetic probes, we explicitly verify this scaling and compute the associated scaling function for the invariant mass spectrum $dN_{\dileptons}/d^2\xT M dM dy_Q$ of pre-equilibrium dileptons. Since the scaling allows for a particularly efficient implementation of pre-equilibrium dilepton production, we further perform state-of-the art phenomenological calculations of (direct) dilepton production in heavy-ion collisions, which demonstrate the importance of event-by-event fluctuations in determining the dilepton yield in the intermediate mass range $3 \GeV < M < 4 \GeV$, where the pre-equilibrium production dominates.

% A recent study has shown that based on such attractors, the pre-equilibrium photon spectrum satisfies a simple universal scaling function in terms of specific shear viscosity $\eta/s$ and entropy density $dS/d\eta \sim {\scriptstyle \left(T\tau^{1/3}\right)_\infty^{3/2}}$~\cite{Garcia-Montero:2023lrd}. In this letter we will present a similar universal scaling function for the invariant mass spectrum of pre-equilibrium dileptons. Based on this we will further compare the yields to other dilepton sources and experimental data.

% Throughout this manuscript, four vectors are denoted as $P$ or $p^\mu$, three vectors as $\pp$, (transverse) two vectors as $\pT$, their magnitude as $p_T$ and energy as $E_p=p^0=|\pp|=p$.

\section{Pre-equilibrium dilepton production} 
Dilepton production in leading order QCD kinetic theory is governed by quark anti-quark annihilation, with the corresponding production rate determined by
\begin{align} \label{eq:DileptonRateDefault}
    &\frac{dN_{\dileptons}}{d^4X d^4Q} = \frac{8\pi}{3} N_c \sum_f q_f^2\alpha_{EM}^2 \int \frac{d^3p_1}{(2\pi)^3 p_1} \frac{d^3p_2}{(2\pi)^3 p_2} \nonumber \\
    &\qquad \times  f_{q_f}(x,P_1) f_{\overline{q}_f}(x,P_2)\delta^{(4)}(Q-P_1-P_2) \, ,
\end{align}
where $Q=P_{l^{+}}+P_{l^{-}}$ is the four momentum of the dilepton pair. Over the short course of the pre-equilibrium phase, the space-time evolution of the non-equilibrium QGP can be locally described in terms of a Bjorken flow, such that the space-time differential is given by $d^4X=\tau d\tau d\zeta d^2\xT$ where $\tau=\sqrt{t^2-z^2}$ is the proper time, $\zeta=\text{arctanh}(z/t)$ is the space-time rapidity and $\xT$ denotes the transverse coordinates.
While in thermal equilibrium, the production rate can be integrated to yield the well known leading order result~(see also \cite{Baier:1988xv,Churchill:2023vpt} for NLO corrections)

\begin{align}
	\frac{dN_{\dileptons}}{d^4Xd^4Q} = \frac{N_C\alpha_{\text{EM}}^2}{12\pi^4}  \sum_f q_f^2 \frac{F(q) }{\exp(q^0/T)-1} \, ,
\end{align}
with 
\begin{equation}
    F(q)=\frac{2T}{q}\ln\bqty{\cosh(\frac{q^0+q}{4T})\left/\cosh(\frac{q^0-q}{4T})\right.}
\end{equation}
the non-equilibrium production of dilepton pairs is governed by 
the phase-space distributions of quarks and anti-quarks $f_{q_{f}}(x,P_1)$ and $f_{\overline{q}_{f}}(x,P_2)$ in Eq.~(\ref{eq:DileptonRateDefault}). 

During the pre-equilibrium phase the QGP is believed to be highly anisotropic in momentum space, as it is subject to a rapid longitudinal expansion, and dominated by gluon degrees a freedom, as these are more copiously produced in the initial (semi-) hard scatterings~\cite{Berges:2020fwq,Iancu:2003xm}. Notably, these effects result in a momentum-space anisotropy~\cite{Kasmaei:2018oag,Kasmaei:2019ofu}, a non-trivial polarization~\cite{Coquet:2023wjk} and an overall suppression of the dilepton yield, as discussed in previous works~\cite{Coquet:2021gms,Coquet:2021lca,Coquet:2023wjk}, where a phenomenological parametrization of the phase-space distributions $f_{q/\bar{q}}$ was employed to compute the pre-equilibrium dilepton production. Conversely, in this work we will compute the pre-equilibrium dilepton production directly from a microscopic calculation in QCD kinetic theory, following the procedure we have presented in  Refs.~\cite{Garcia-Montero:2023lrd}. However, before we present results of our microscopic calculations, it proves particularly insightful to perform a dimensional scaling analysis of the pre-equilibrium dilepton yield.

\subsection{Scaling of pre-equilibrium dilepton production} Generally, the pre-equilibrium production of electromagnetic probes will proceed differently at different positions $\xT$ in the transverse plane and depend on the timescale $\tau_{\rm eq}(\xT)$, and the temperature $T_{\rm eq}(\xT)$ at which the system locally approaches a viscous hydrodynamic description. By dimensional analysis, it is natural to assume, that the local production rate  $ \frac{dN_{\dileptons}}{d^4X d^4Q}$ at each transverse position $\xT$ only depends on the dimensionless ratios $\tau/\tau_{\rm eq}(\xT)$ and $Q/T_{\rm eq}(\xT)$, 
\begin{align}
    \frac{dN_{\dileptons}}{d^4X d^4Q} = \frac{dN_{\dileptons}}{d^4X d^4Q} \left( \frac{\tau}{\tau_{\rm eq}(\xT)},\frac{Q}{T_{\rm eq}(\xT)} \right)
\end{align}
such that upon integrating over the production time $\tau$, the space-time time rapidity $\zeta$, and the transverse momentum $q_T$ of the dilepton pair, the local yield $\frac{dN_{\dileptons}}{d^2\xT M dM dy_{Q}} = \int d\tau \tau \int d\zeta \int d^2q_T \frac{dN_{\dileptons}}{d^4X d^4Q}$ per unit rapidity $y_{Q}$ and invariant mass $M=\sqrt{Q^2}$ is determined by
\begin{align}\label{eq:SpectrumScaling}
\frac{dN_{\dileptons}}{d^2\xT M dM dy_{Q}} = \tau_{\rm eq}^2(\xT) T_{\rm eq}^{2}(\xT)~\mathcal{N}_{{\dileptons}}\left(\frac{M}{T_{\rm eq}}\right)
\end{align}
where $N_{l^{+}l^{-}}$ is a scaling function, which only depends on the ratio $M/T_{\rm eq}$, and is determined by
\begin{align}
\mathcal{N}_{{\dileptons}}\left(\frac{M}{T_{\rm eq}}\right)&=\int d\frac{\tau}{\tau_{\rm eq}} \frac{\tau}{\tau_{\rm eq}} \int d\zeta \int d^2\frac{q_T}{T_{\rm eq}} \\
&\quad \frac{dN_{\dileptons}}{d^4X d^4Q} \left( \frac{\tau}{\tau_{\rm eq}},\frac{Q}{T_{\rm eq}} \right)
\end{align}
While the calculation of the scaling function $\mathcal{N}_{\dileptons}\left(\frac{M}{T_{\rm eq}}\right)$ is highly non-trivial, it is comparatively straightforward to isolate the dependence on the local entropy density and coupling strength of the system. We first note that different studies at weak~\cite{Kurkela:2018oqw,Du:2020dvp} and strong coupling~\cite{Florkowski:2017olj,Romatschke:2017ejr,Soloviev:2021lhs}) indicate, that equilibration of the out-of-equilibrium QGP is controlled by the equilibrium relaxation rate $\Gamma_{\rm eq}=T_{\rm eq}/(4\pi \eta/s)$, such that hydrodynamics becomes applicable when the dimensionless scaling variable $\tilde{w} = \Gamma_{\rm eq}\tau_{\rm eq}$
is of order unity, such that 
\begin{align}
\label{eq:ThermalizationRel}
    \tau_{\rm eq} T_{\rm eq} \approx 4\pi \eta/s \, .
\end{align}
Since the equilibrium temperature $T_{\rm eq}$ can be related to the local entropy density of the system $\frac{dS}{d^2\xT d\zeta}$, which in turn is directly related to the multiplicity of charged particles produced in the collisions, one further obtains the relation
\begin{align}
\label{eq:EntropyRel}
    \frac{2 \pi^2}{45} \nu_{\rm eff} \tau_{\rm eq} T_{\rm eq}^3 \approx \frac{dS}{d^2\xT d\zeta}
\end{align}
where $\nu_{\rm eff}\approx 32$ denotes the effective number of bosonic degrees of freedom~\cite{Borsanyi:2013bia,HotQCD:2014kol}. Denoting the dependence on the local entropy density in terms of the quantity 
$(T\tau^{\frac{1}{3}})^{3/2}_{\infty}=\sqrt{\frac{dS}{d^2\xT d\zeta} / \left(\frac{2 \pi^2}{45} \nu_{\rm eff}\right)}$, by virtue of Eqns.~(\ref{eq:ThermalizationRel}) and (\ref{eq:EntropyRel}) the equilibration time and equilibration temperature can then be determined as
\begin{align}
    \tau_{\rm eq} &\approx (T\tau^{\frac{1}{3}})^{-3/2}_{\infty} \left( 4\pi \eta/s \right)^{3/2} \\
    T_{\rm eq} &\approx (T\tau^{\frac{1}{3}})^{3/2}_{\infty}~ \left( 4\pi \eta/s \right)^{-1/2}
\end{align}
and the pre-equilibrium dilepton yield can be expressed as
\begin{align}\label{eq:ScalingFormula}
\frac{dN_{\dileptons}}{d^2\xT M dM dy_{Q}} = (4\pi \eta/s)^2~\mathcal{N}_{\dileptons}\left( \frac{\sqrt{4\pi \eta/s}~M}{ (T\tau^{1/3})^{3/2}_{\infty}}\right)
\end{align}
Strikingly, the above scaling can be derived on rather general grounds, and thus easily extended to a variety of other quantities, such as e.g. the transverse momentum ($q_T$) spectrum, which takes the form
\begin{align}
&\frac{dN_{\dileptons}}{d^2\xT d^2 \q_T M dM dy_{Q}} = (4\pi \eta/s)^3~(T\tau^{\frac{1}{3}})^{-3}_{\infty} \\
& \quad \mathcal{N}_{\dileptons}\left( \frac{\sqrt{4\pi \eta/s}~M}{(T\tau^{\frac{1}{3}})^{3/2}_{\infty}}, \frac{\sqrt{4\pi \eta/s}~q_T}{(T\tau^{\frac{1}{3}})^{3/2}_{\infty}}\right)
\end{align}
Most importantly, such scaling can also be expected to be present for other processes, such as e.g. the pre-equilibrium production of photons~\cite{Garcia-Montero:2023lrd}, heavy flavor quarks~\cite{Du:2023izb} or pre-equilibrium jet quenching~\cite{Andres:2022bql}, which will greatly facilitate their phenomenological treatment in the future.

\subsection{Dilepton production from non-equilibrium QGP evolution.}
\begin{figure*}[t!]
\begin{center}
\centering
\begin{subfigure}[b]{0.48\textwidth}
\includegraphics[width=0.94\textwidth]{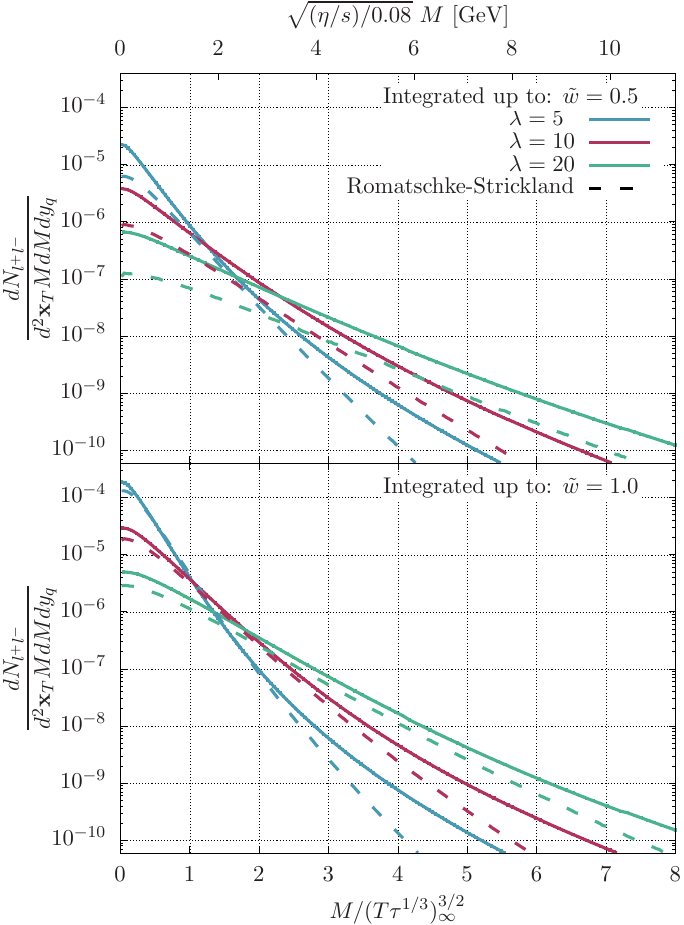}
\caption{Dilepton spectra}\label{fig:MSpectruma}
\end{subfigure}
\begin{subfigure}[b]{0.48\textwidth}
\includegraphics[width=0.94\textwidth]{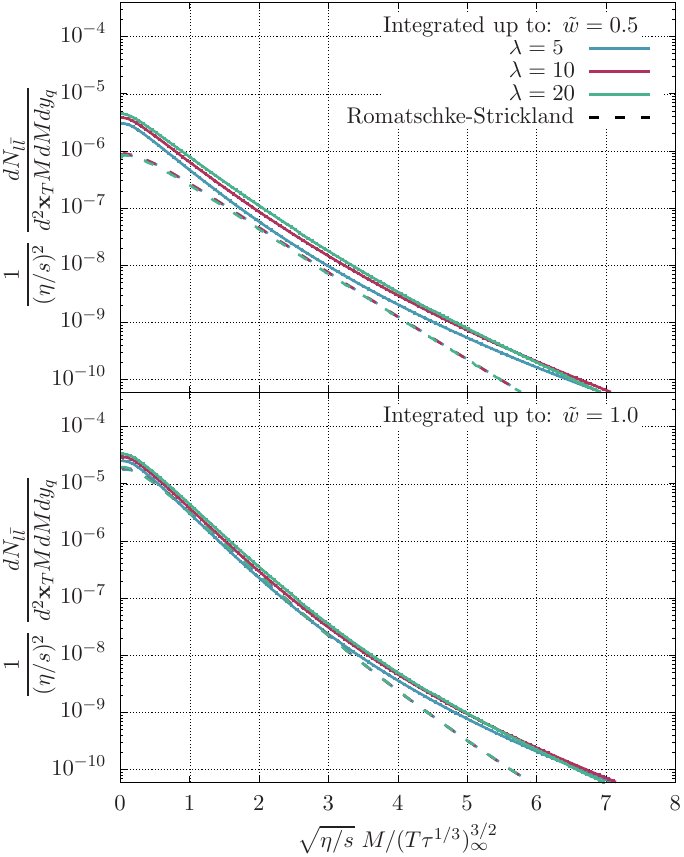}
\caption{Universal scaling curve}\label{fig:MSpectrumb}
\end{subfigure}
\caption{\label{fig:MSpectrum} Spectrum of dileptons produced during the pre-equilibrium stage as a function of (a) $M/(T\tau^{1/3})_\infty^{3/2}$ and (b) $\sqrt{\eta/s} M/(T\tau^{1/3})_\infty^{3/2}$. Different colors correspond to different coupling strengths $\lambda=5,10,20$ and different panels correspond to different final integration times $\tilde w=0.5,1.0$. Spectra in the right panels are divided by $(\eta/s)^2$ in order to obtain the universal scaling curve.}
\end{center}
\end{figure*}

In order to verify the predicted scaling we compute the pre-equilibrium spectrum of dileptons for different values of the coupling $\lambda= 4\pi N_c \alpha_s$ using a QCD kinetic theory description, where we incorporate all leading order processes of quarks and gluons into the Boltzmann equation to compute the respective distribution functions during the pre-equilibrium evolution of the QGP~\cite{Arnold:2002zm,Garcia-Montero:2023lrd,Du:2020dvp}. Within this framework infrared divergencies are regulated by an isotropic screening~\cite{Kurkela:2015qoa,Kurkela:2018vqr,Kurkela:2014tea,Kurkela:2018oqw,Du:2020dvp} and in-medium effects associated with the Landau-Pomeranchuk-Migdal (LPM) effect are treated via an effective vertex resummation. Starting from a gluon dominated initial state as in~\cite{Garcia-Montero:2023lrd}, we numerically solve the Boltzmann equation to determine the evolution of the phase-space distributions $f_{q/\overline{q}}$, which then enables us to compute the dilepton rate at each time step. The procedure for the evolution of the pre-equilibrium QGP, which serves as a background to the dilepton production rate,\cref{eq:DileptonRateDefault}, was described in Ref.~\cite{Garcia-Montero:2023lrd}. For the numerical computation of the dilepton rate we use the same setup as in the aforementioned photon production study, including the same initial conditions and the assumption of boost invariance.

Since the QGP created in relativistic heavy-ion collisions at RHIC and LHC energies remains strongly coupled, as indicated by a small shear-viscosity to entropy density ratio $\eta/s$, the framework of QCD Kinetic Theory is not directly applicable and the results need to be extrapolated to realistic coupling strength for phenomenological application. In this context, it is important to point out that studies of the pre-equilibrium evolution in various weakly~\cite{Kurkela:2015qoa,Kurkela:2018wud,Giacalone:2019ldn,Du:2020zqg} and strongly coupled theories~\cite{Romatschke:2017vte,Beuf:2009cx} indicate that the processes of kinetic and chemical equilibration are governed by a single timescale related to the equilibrium relaxation time  $\tau_R$, determined by the shear viscosity to entropy density ratio~\cite{Schlichting:2024uok}. While we use weakly-coupled QCD Kinetic Theory to calculate the pre-equilibrium evolution of the phase-space distributions of quarks and gluons, we therefore anticipate that the non-equilibrium effects of the bulk anisotropy and gluon dominance on dilepton production can be considered representative of different microscopic models. Therefore, we also anticipate that the results can be extrapolated to viable coupling strength based on the scaling variables $\tilde{w} \approx \tau T_{\rm eff}(\tau)/4\pi\eta/s$, and ${(\eta/s)^{1/2}} M/{(\tau^{1/3} T)^{3/2}_\infty}$; showing weak dependence on coupling strength and consistent behavior across different microscopic models.

\begin{table}[t]
    \centering
    \begin{tabular}{c|c|c|c}
        $\lambda$ & 5 & 10 & 20 \\
        \hline
        $\alpha_s$ & 0.133 & 0.265 & 0.531 \\
        \hline
        %$\eta/s$ & 2.71614 & 0.994138 & 0.384624 \\
        $\eta/s$ & 2.716 & 0.994 & 0.385 \\
        \hline
        %$(\tau T^3)_\infty/Q_s^2$ & 0.799219 & 0.611515 & 0.473776 \\
        $(\tau^{1/3} T)_\infty/Q_s^{2/3}$ & 0.799 & 0.612 & 0.474 \\
        %\hline
        %$\tilde{w}_{\text{min}}$ & 0.015328 & 0.0352155 & 0.0765397
        %$\tilde{w}_{\text{min}}$ & 0.015 & 0.035 & 0.077
    \end{tabular}
    \caption{Strong coupling $\alpha_s$, shear viscosity to entropy density ratio $\eta/s$ and asymptotic temperature $(\tau^{1/3} T)_{\rm \infty}/Q_s^2$ for different coupling strengths $\lambda$.}
    \label{tab:Constants}
\end{table}

Numerical results for the pre-equilibrium spectrum of dileptons as a function of invariant mass are presented in \cref{fig:MSpectruma}, where we show the integrated dilepton yield up to $\wTilde=0.5$ (top), during the thermalization process of the QGP and $1.0$ (bottom), when visc. hydrodynamics becomes applicable. When comparing the results for different coupling strength $\lambda=5,10,20$, which correspond to different values of the specific shear viscosity $\eta/s$, we express the spectra in terms of the invariant mass 
$M/{(\tau^{1/3} T)^{3/2}_\infty}$ (in \cref{fig:MSpectruma}) and the
re-scaled invariant mass variable ${(\eta/s)^{1/2}} M/{(\tau^{1/3} T)^{3/2}_\infty}$  (in \cref{fig:MSpectrumb}). The specific shear viscosity as well as the asymptotic constant $(\tau^{1/3} T)_\infty$ are obtained by comparison to the first order viscous hydrodynamic expansions of the longitudinal pressure resp. the effective temperature (see~\cite{Kurkela:2018vqr,Du:2022bel,Garcia-Montero:2023lrd} for more details). The results for $\lambda=5,\, 10,\, 20$ are compactly summarized in \cref{tab:Constants}.

%It is very important to note that we use the QCD KT to describe the dilepton production as a way to encompass the microscopic and out-of-equilibrium aspects of  the early time evolution. However, what for us is the background evolution, the early pre-equilibrium stage,  does not directly depend on QCD KT to describe  the pre-equilibrium dynamics of the QGP. In the last years, studies of isotropization and hydrodynamic behavior in various weakly~\cite{Kurkela:2015qoa,Kurkela:2018wud,Giacalone:2019ldn,Du:2020zqg} and strongly coupled theories~\cite{Romatschke:2017vte,Beuf:2009cx} indicate that this process is governed by a single timescale related to the equilibrium relaxation time  $\tau_R$, determined by the shear viscosity to entropy density ratio~\cite{Schlichting:2024uok}. While we use a parametrization from QCD Kinetic Theory to calculate the effects of non-equilibrium evolution, our results are broadly applicable across different microscopic models.

%Another important point is that  QCD KT is at the edge of its applicability for determining the correct  $\eta/s$ value. However, recent findings suggest that results from various microscopic theories can be extrapolated to viable $\eta/s$ values based on the scaling variable $\tilde{w} \approx \tau T_{\rm eff}(\tau)/4\pi\eta/s$ showing weak dependence on coupling strength and consistent behavior across models. We observe this behavior too in Fig.~\ref{fig:MSpectruma} and ~\ref{fig:MSpectrumb}. 
Thus, variations in the coupling reflect the sensitivity of this scaling parameter, as illustrated in Fig.~\ref{fig:MSpectrum}.

When comparing the results for different coupling strength as a function of $M/{(\tau^{1/3} T)^{3/2}_\infty}$, we see that stronger couplings tend to produce more large-$M$ dileptons. Setting $(\tau^{1/3} T)^{3/2}_\infty=0.402\GeV$ as in~\cite{Garcia-Montero:2023lrd}, to mimic the conditions in central Pb+Pb collisions at LHC energies, this occurs for invariant masses $M \gtrsim 2\GeV$. Below $M \lesssim 2\GeV$ this changes and more weakly coupled systems produce more dileptons during the pre-equilibrium phase, i.e. for $\tilde{w} <1$. We further see that high-$M$ dileptons are predominantly produced at the earliest times of the evolution. Indeed, the curves for $\wTilde=0.5$ are almost not modified during the evolution up to $\wTilde=1.0$. However, in the mean time the QGP produces many low-mass dileptons, such that the spectrum increases significantly in the low-$M$ regime.

Besides the QCD kinetic theory  results, we also show results for an effective parameterization (EP) of the distribution functions already used in previous works for pre-equilibrium dilepton production~\cite{Coquet:2021gms,Coquet:2021lca,Coquet:2023wjk}, where the authors used anisotropic equilibrium distribution functions of the Romatschke-Strickland form (see Fig~\ref{fig:MSpectrumb} and Ref.~\cite{Romatschke:2003ms}), together with an overall quark suppression factor. Compared to the QCD KT results, the spectra computed from the effective parameterization are suppressed at both times, although the suppression is stronger for earlier times. This is expected as the description of pre-equilibrium dynamics in terms of anisotropic equilibrium distributions is worse for earlier time dynamics. 
%In fact, the quark suppression is zero for $\wTilde \leq 0.2$, such that dileptons before that time are not included in the EP-spectra. 
Over the course of the evolution, the spectra for the effective parametrization get closer to the QCD KT results and for smaller couplings we find almost the same spectra for the low and intermediate mass range.

To verify the universal scaling, we re-plot the same spectra in \cref{fig:MSpectrumb} with the predicted scaling from \cref{eq:ScalingFormula}. By construction, the spectra obtained using the effective parametrization perfectly collapse to each other, while the QCD KT results also get much closer to each other. Since the scaling function $N_{l^{+}l^{-}}$ is derived on basis of an attractor solution, the scaling for the QCD KT results works better and better for later evolution. Especially for $\wTilde \gtrsim 1$, which is in accordance with the hydrodynamization time found in earlier studies in the context of QCD kinetic theory~\cite{Giacalone:2019ldn,Du:2020zqg,Kurkela:2018vqr,Kurkela:2018wud}, the scaling works very well, whereas at very early times the spectra are sensitive to the initial conditions for the kinetic evolution.
We discuss the residual deviations from scaling further in \cref{app:dependence}.

\section{Phenomenology of the pre-equilibrium dilepton production}
One remarkable consequence of the scaling formula is the practical usage for phenomenological calculations and comparisons to experimental data. In order to demonstrate this, we match the QCD KT evolution to the initial conditions of the subsequent hydrodynamic evolution  of the QGP. By use of the scaling formula in Eq.~\ref{eq:ScalingFormula}, the only other ingredients to compute the spectrum of dileptons produced during the pre-equilibrium are the (local) temperature $T(\tau_\hydro,\xT)\equiv T_\hydro(\xT)$ at initialization time $\tau_\hydro$ and the specific shear-viscosity $\eta/s$. Based on this information, the local values of the scaling variable $\tilde{w}(\xT)$ can then be determined as
\begin{equation}
    \tilde{w}(\xT) = \frac{\tau_{\rm hydro} T_{\text{hydro}}(\xT)}{4\pi \eta/s}
   \label{eq:hydro_wtilde} \, ,
\end{equation}
and similarly, the energy scale $(\tau^{1/3}T)_{\infty}(\xT)$ can be obtained according to~\cite{Kurkela:2018vqr,Du:2022bel}
\begin{equation}
   (\tau^{1/3}T)_{\infty}(\xT) = \tau_{\rm hydro}^{1/3}\,\left(T_{\rm hydro}(\xT) + \frac{2}{3} \frac{\eta/s}{\tau_{\rm hydro}}\right) \, 
\end{equation}
%\begin{figure}
 %   \centering
%\includegraphics[width=1.0\linewidth]%{FiguresPheno/UniversalTimeDependence_eta_s_0.16_thydro_1fm.pdf}
  %  \caption{Scaling time evolution of the dilepton spectra for $0-5\%$ centrality Pb-%Pb collisions at $\sqrt{s_{NN}} = 5.02$~TeV. The evolution is presented for %diverse scaling time-slices.}
%    \label{fig:pheno_break_down}
%\end{figure}
accounting for first order viscous corrections. 
Once those two quantities are determined at each point in the transverse plane, the total pre-equilibrium dilepton yield is then given by
\begin{align}
    \frac{dN_{\dileptons}}{MdM dy_Q} = (4 \pi \eta/s)^2 \! \! \int \! d^2\xT ~\mathcal{N}_{\dileptons} \pqty{\tilde{w}(\xT), \textstyle \frac{\sqrt{\eta/s} ~M}{\left(T\tau^{1/3}\right)_\infty^{3/2}(\xT)}} \, .
\label{eq:phenomastereq}
\end{align}
We compute the event-by-event dilepton spectra at midrapidity for Pb-Pb collisions at $\sqrt{s_{NN}} = 5.02$~TeV. To account for a background evolution well fit to data, we created the events using the hybrid model \emph{Trajectum} ~\cite{Nijs:2020ors}, and used one of the consistent parameter sets extracted using the Bayesian analysis in  Ref.~\cite{Giacalone:2023cet}. We focus on the $0-5\%$ class, and obtain the pre-equilibrium dilepton yield by integrating event-by-event over the initial profiles for the hydrodynamical evolution as in~\cref{eq:phenomastereq}. By matching the initial profiles to the subsequent hydrodynamic evolution at $\tau_{hydro}=1.0$~fm as in~\cite{Garcia-Montero:2023lrd}, the thermal QGP contribution is then found by folding the thermal dilepton production rate with the background hydrodynamic evolution from $\tau_{hydro}=1.0$~fm onwards\footnote{The QGP rate is folded with the hydro evolution for temperatures higher than $T_c=154$~MeV. Below this temperature, hadronic degrees of freedom dominate, and the rate can be computed from effective theories~\cite{Nijs:2020ors}. Nevertheless, the hadronic rates are too soft to contribute significantly to the invariant mass range relevant to this work.}. Clearly, one of the main advantages of using this description is that it guarantees for a smooth matching between pre-equilibrium and hydrodynamical dilepton production, such that the overall spectrum of in-medium dileptons is rather insensitive to the matching time $\tau_{\rm hydro}$, as discussed further in \cref{app:dependence}.

%In \cref{fig:pheno_break_down}, we have broken the event-averaged pre-equilibrium dilepton spectrum into several $\wTilde$ dependent contributions, specifically the ranges $0.1-0.2$, $0.2-0.4$, $0.4-0.6$ and $0.6-0.8$. This is to illustrate how dilepton pairs radiated from different stages of the pre-equilibrium evolution behave. For the ranges with higher $\wTilde$, one can see that the spectra obtained are increasingly softer, while earlier times yield harder spectra.

\begin{figure}
	\centering
	\includegraphics[width=1.0\linewidth]{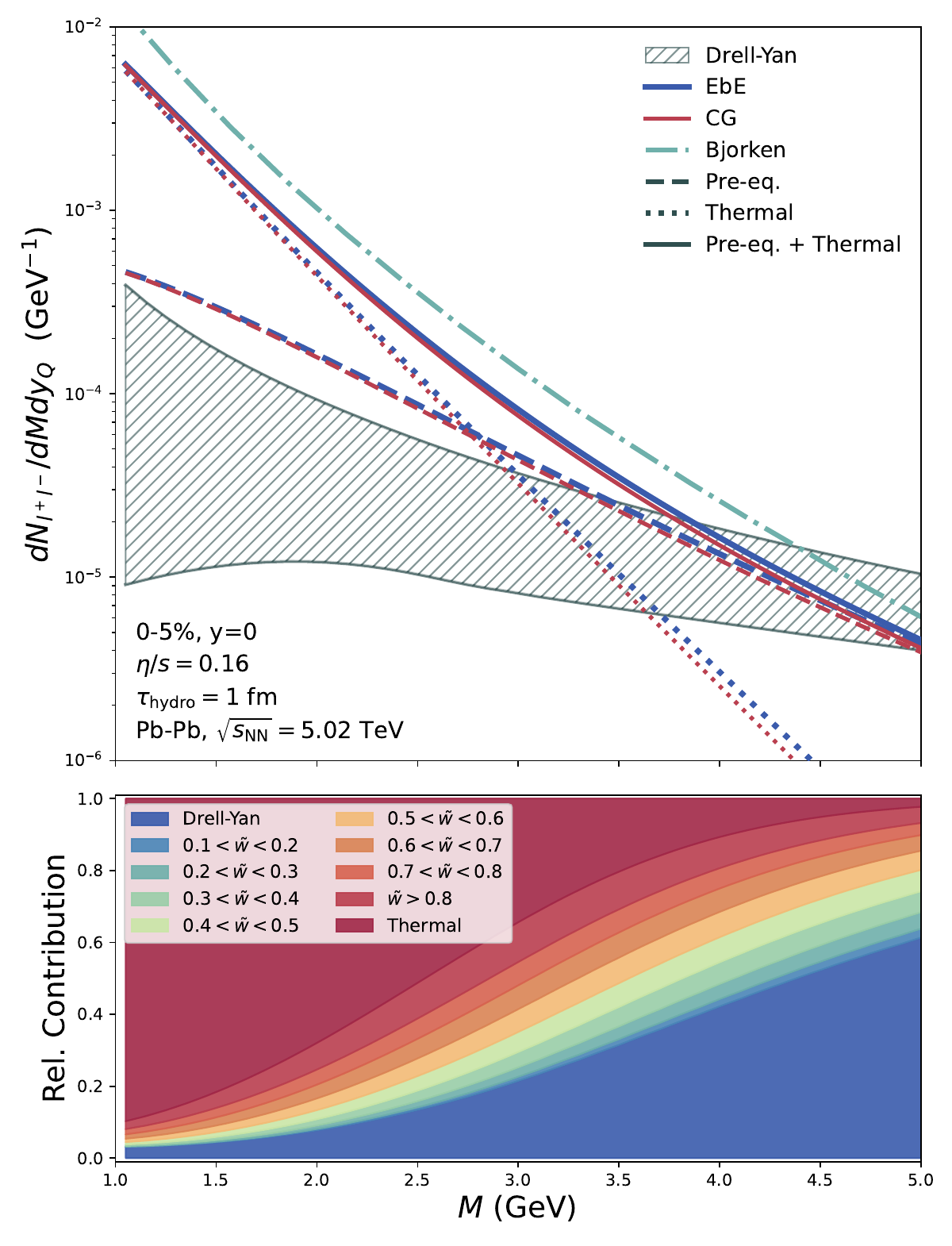}
	\caption{ (\textit{Upper panel}) Comparison of the in-medium dilepton spectrum for $0-5\%$ centrality Pb-Pb collisions at $\sqrt{s_{NN}} = 5.02$~TeV to their corresponding Drell-Yan contribution. The in-medium contributions are shown for three different cases (transversely homogeneous Bjorken expansion (data taken from~\cite{Coquet:2021lca}), realistic EbE ICs, and smooth initial conditions, the average of the latter). (\textit{Lower panel}) Relative contribution of the different contributions to the total yield (in-medium + Drell-Yan). The EbE pre-equilibrium has been split into different $\tilde{w}$ ranges for better comparison.}
	\label{fig:medium_vs_DY}
\end{figure}

We present our main phenomenological result in \cref{fig:medium_vs_DY}, where we compare the in-medium dilepton production, to the Drell-Yan (DY) contribution, computed to NLO precision with the EPPS nPDFs~\cite{Eskola:2016oht} by using the DYTurbo software~\cite{Camarda:2019zyx}, where as in previous works~\cite{Coquet:2021lca} the centrality dependence of the nPDFs was neglected and $T_{AA}$ scaling was assumed. 
While the top panel of \cref{fig:medium_vs_DY} shows the total dilepton yield $dN_{\dileptons}/dMdy_Q$, the lower panel shows a break-up of the total yield into DY and thermal QGP contributions, as well as in the the pre-equilibrium contribution from ranges of evolution time $\tilde{w}$, and clearly illustrates how larger invariant masses correspond to earlier production times. 

While at higher invariant mass ranges the DY contribution is well constrained and therefore presents small uncertainties, at low $M$s, it presents large error bands, due to large scale uncertainties.  
Despite these large theoretical uncertainties, we find that the pre-equilibrium dilepton contribution obtained from realistic event-by-event QCD KT+Hydro simulations dominates over both the DY and thermal QGP background for invariant masses $\sim 3\GeV$. 

Beyond the results obtained from event-by-event (EByE) simulations, we have also computed the in-medium dilepton spectra for a coarse grained (CG) description, where similar to ~\cite{Endres:2014zua,Endres:2016isp} we compute the emission for a single average event, as well as comparing to previous obtained results for a simple Bjorken scaling scenario~\cite{Coquet:2021lca}, where transverse dynamics of the QGP is completely neglected. While the Bjorken scaling underestimates the cooling of the QGP, and thus overestimates the dilepton yield by almost a factor of three, the coarse grained averaged description provides a very good description for low invariant masses $\lesssim 2\GeV$. Nevertheless, the coarse grained description also but underestimates the high $M$ tails of in-medium dilepton production, which are more likely produced from local hot spots, and we therefore conclude from this analysis, that an event-by-event description is desirable to provide accurate predictions for high and intermediate mass dileptons. This difference between the EbE and CG descriptions becomes more important for less central collisions. 

We close our discussion by putting our calculations of pre-equilibrium dilepton production in the context of other recent works, discussing the most important similiarities and differences in methodology and findings. While our studies are based on event-by-event calculations with microscopic non-equilibrium rates, a number of works have addressed the same question of using dileptons as probes to extract information about  the early stages, following the interest to address the same question using real photons \cite{Berges:2017eom,Monnai:2019vup,Garcia-Montero:2019kjk,Garcia-Montero:2019vju,Garcia-Montero:2023lrd}.

In~\cite{Coquet:2021lca,Coquet:2021gms}, the authors estimate the relative contributions of the in-medium production (pre-equilibrium and hydro stages) with respect to the expected Drell-Yan contribution and find that there is a window $M\approx 2-3.5$ GeV in which the pre-equilibrium dominates. In this estimate, the authors consider a  transversely homogeneous system undergoing a simple Bjorken expansion (as in our transversely homogenous scenario),  and the non-equilibrium dilepton rates are determined from a QCD KT inspired parametrization of the phase-space distribution of quarks and gluons (as in our Romatsche-Strickland paramtrization). Even though the Bjorken scenario may lead to a slight overestimation of the pre-equilibrium dilepton rate, the results of  \cite{Coquet:2021lca,Coquet:2021gms} are generally consistent with our findings; moreover the authors also  proposed the polarization of dilepton pairs as a possible means to discriminate between pre-equilibrium and Drell-Yan contributions~\cite{Coquet:2023wjk} and it would be interesting to extend our methodology to such polarization observables.

In~\cite{Wu:2024pba} the authors perform event-by-event studies of pre-equilibrium dilepton production. Based on the macroscopic pre-equilibrium evolution of the energy-momentum tensor in KøMPøST~\cite{Kurkela:2018vqr,Kurkela:2018wud}, the authors extract an effective temperature, which is then employed in thermal rates to compute the pre-equilibrium production. Non-equilibrium effects on the rates are accounted for by introducing a phenomenological quark suppression factor, inspired by Refs.~\cite{Kurkela:2018xxd,Kurkela:2018oqw}. However, the authors do not take into account the modified shapes of the pre-equilbrium momentum distributions of quarks and gluons, (see. e.g.~\cite{Garcia-Montero:2023lrd}), which is particularly relevant for the hardest part of the spectrum produced at the earliest times. Qualitatively, the results of ~\cite{Wu:2024pba} give a very similar picture for the expected mass spectrum of dileptons in ultra-relativistic HICs, albeit the calculation of  ~\cite{Wu:2024pba} slighty overshoots our results. This gives a wider window of relevance for the measurement of pre-equilibrium dileptons. 

%Recently, a number of works have addressed the question of using dileptons as probes to extract information of the early stages, following the interest to address the same question using real photons \cite{Berges:2017eom,Monnai:2019vup,Garcia-Montero:2019kjk,Garcia-Montero:2019vju,Garcia-Montero:2023lrd}. In Refs.~\cite{Coquet:2021lca,Coquet:2021gms}, the authors make an early estimation of the relative contributions of the in-medium production (pre-equilibrium and hydro stages) with respect to the extpected Drell-Yan contribution and find that there is a window ($M\approx 2-3.5$ GeV) in which the pre-equilibrium dominates. This estimation was produced using a Bjorken expanding, transversely homogeneous system. While the pre-equilibrium was performed using  a general Ansatz for the scaling of the distributions (see parametrization in Ref.~\cite{Romatschke:2003ms}), the hydrodynamical stage is modeled with Bjorken expansion, i.e. ignoring transverse cooling. Using this methodology, the authors proposed also using the polarization of dilepton pairs to the contributions of DTY and in-medium production~\cite{Coquet:2023wjk}. 

%Essentially, the work here presented builds upon these estimations by abandoning some of the simplifying assumptions. Apart from including a more realistic pre-equilibrium evolution, which accounts for the initial quark suppression naturally, we add realistic EbE profiles followed by a state-of-the-art (and thoroughly benchmarked) hydrodynamical evolution, and subsequent particlization and afterburner. 

\section{Summary and Outlook.}
We computed the pre-equilibrium spectrum of dileptons in QCD kinetic theory and showed that it exhibits a simple universal scaling in terms of the specific shear viscosity $\eta/s$ and entropy density $(T\tau^{{1}/{3}})_{\infty}$. Such scaling functions are powerful tools as they allow for event-by-event calculations of the corresponding spectra by matching $\eta/s$ and $(T\tau^{{1}/{3}})_{\infty}$ to realistic values of a heavy-ion collision. By performing realistic event-by-event simulations of pre-equilibrium, thermal QGP and Drell-Yan dilepton production for $\sqrt{s_{NN}} = 5.02$~TeV Pb-Pb collisions, we find that the pre-equilibrium production dominates over the irreducible background for invariant masses $\sim 3\GeV$. However, this result only serve as an illustrative example and in order to make full use of our results for phenomenological applications, we have included the pre-equilibrium scaling functions into the initial state framework \kompost. This version including photons and dileptons is called Shiny\kompost{} and is publicly available under~\cite{KoMPoST}.

Since the pre-equilibrium scaling in Eq.~\eqref{eq:ScalingFormula} can be derived on rather general grounds, it can also be extended to fully differential dilepton observables, which is highly beneficial for the development of a Monte-Carlo generator to facilitate the detection of pre-equilibrium dileptons in present and future heavy-ion experiments. Since earlier studies showed that the polarization of dileptons is a key observable to distinguish the pre-equilibrium emission from other sources~\cite{Coquet:2023wjk}, clearly a next step will be to derive analogous scaling functions for polarization observables. However, the use of pre-equilibrium scaling functions is by no means limited to electromagnetic probes, an extension to relevant topics such as heavy-flavor production~\cite{Uphoff:2010sh,Zhang:2007yoa,Zhou:2016wbo,Du:2023izb}, heavy-quark diffusion~\cite{Boguslavski:2020tqz} or jet energy loss~\cite{Zhou:2024ysb} would be interesting to see and deserves further theoretical attention.

\section*{Acknowledgements}
We thank Maurice Coquet, Travis Dore, Xiaojian Du, Stephan Ochsenfeld, Jean-Yves Ollitrault, Michael Winn for valuable discussions. The authors want to thank Aleksas Mazeliauskas for collaboration on related work. OGM, PP and SS acknowledge support by the Deutsche Forschungsgemeinschaft (DFG, German Research Foundation) through the CRC-TR 211 ‘Strong-interaction matter under extreme conditions’-project number 315477589 – TRR 211. OGM and SS acknowledge also support by the German Bundesministerium für Bildung und Forschung (BMBF) through Grant No. 05P21PBCAA. The authors acknowledge computing time provided by the Paderborn Center for Parallel Computing (PC2) and the National Energy Research Scientific Computing Center, a DOE Office of Science User Facility supported by the Office of Science of the U.S. Department of Energy under Contract No. DE-AC02-05CH11231.

\appendix

\section{Sensitivity of pre-equilibirum dilepton production on coupling constant $(\lambda)$ and matching time $(\tau_{\rm Hyhdro}$)}

Below we discuss the dependence of the scaling function on the coupling constant $\lambda$, and the sensitivity of the integrated dilepton yield on the matching time, $\tau_{\rm hydro}$.

\label{app:dependence}
\begin{figure}[t]
    \centering
    \includegraphics[width=0.95\linewidth]{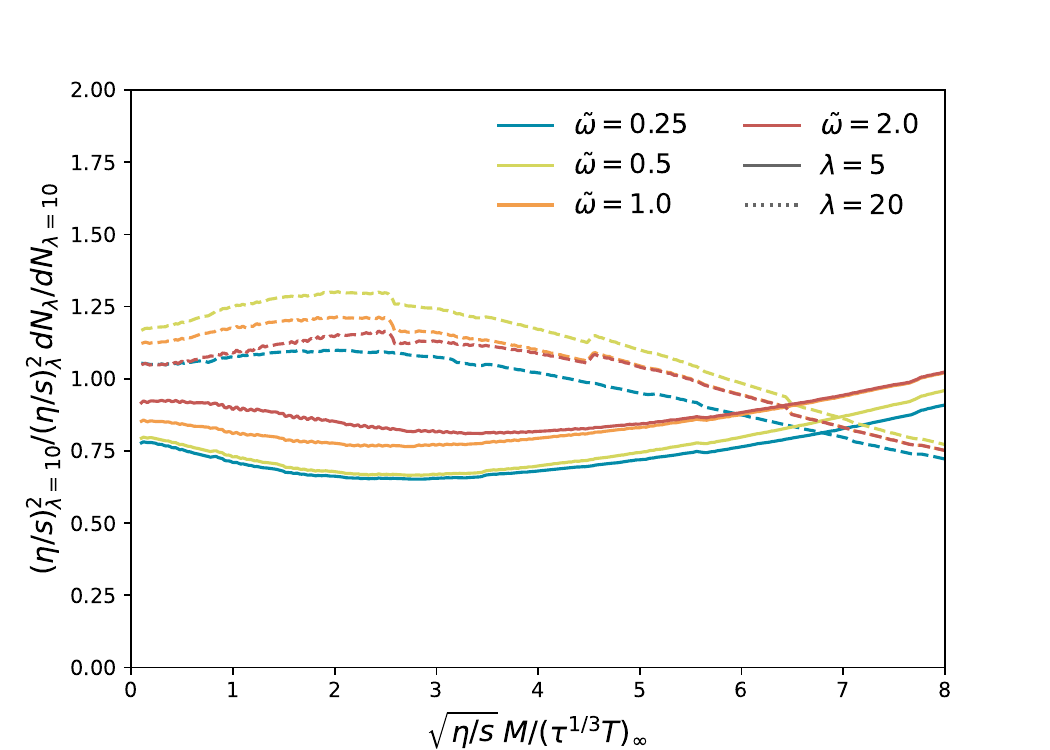}
    \caption{Scaling function comparison for different coupling constants ($\lambda = 5, 10, 20$), normalized to $\lambda = 10$. Deviations remain within 30–35\% for relevant mass ranges and decrease to $<20\%$ for $\tilde{w} \geq 1$, supporting the robustness of the scaling approach.}
    \label{fig:lambda dependence}
\end{figure}

\subsection{Sensitivity ot the coupling constant, $\lambda$}

By virtue of the scaling functions, the leading coupling dependence of the integrated yields is taken into account. While, by construction, this scaling is exact for the Romatschke-Strickland form of the phase-space distributions, the scaling relations \cref{eq:ScalingFormula} are only approximate for EKT, and it is  thus important to assess residual deviations from scaling 

%valid for regions of low and intermediate $M$ and large $\tilde{\omega}$.}

In \cref{fig:lambda dependence}, we have included a comparison of the resulting scaling function for the choice of three different coupling constants. This is presentad as the ratio of the extremal values, $\lambda=(5,20)$, normalized by the scaling function for the choice of $\lambda=10$.
For the phenomenologically relevant values of the rescaled invariant mass, the scaling holds down to deviations no larger than $30-35\%$. However, it is important to emphasize that for large $\tilde{\omega}$, where the phase-space distributions in EKT become more and more similar to the Romatsche-Strickland form, 
deviations get smaller and smaller, such that already for $\tilde{\omega}=1$ we find that deviations are no larger than $20\%$.

Since the phenomenological application of the scaling formula only makes sense in the cases where the kinetic system has already effectively thermalized, the switching  to thermal production has to be performed around $\tilde{\omega} \gtrapprox 1$, we consider this level of accuracy sufficient for phenomenological applications.

\subsection{Sensitivity to the matching time $\tau_{\rm hydro }$}

\begin{figure}[t!]
    \centering
    \includegraphics[width=0.99\linewidth]{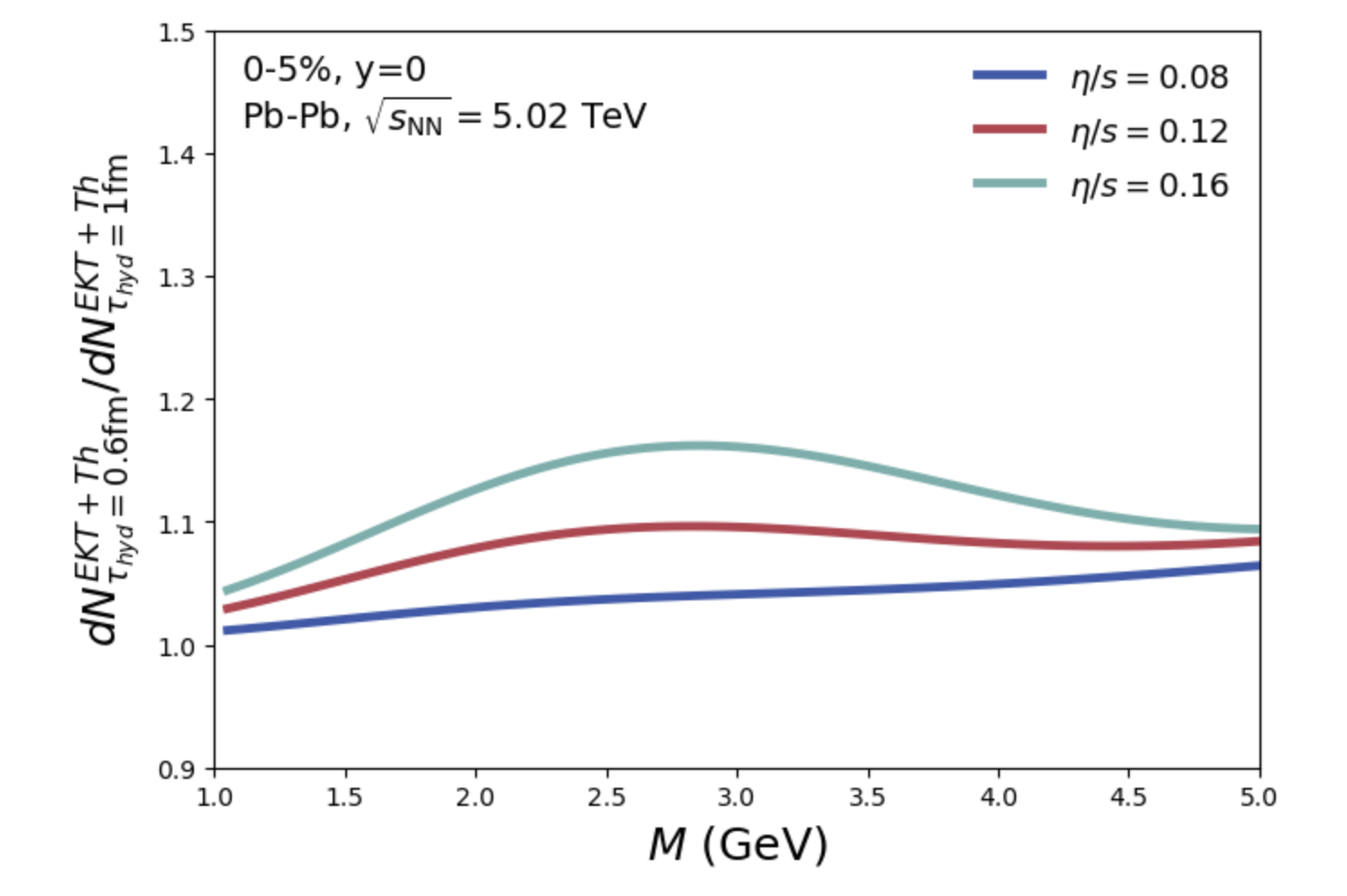}
    \caption{Effect of hydrodynamization time $\tau_{\text{hydro}}$ on dilepton production. The ratio of spectra for $\tau_{\text{hydro}} = 0.6$ and 1.0 fm shows up to 20\% variation, with larger deviations for higher shear viscosity.}
    \label{fig:hydro-dependenced}
\end{figure}

By following our procedure in Ref.~\cite{Garcia-Montero:2023lrd}, we have studied how changes in the switching time, $\tau_{\rm hydro}$, affect the overall invariant mass spectrum of dileptons. In a idealized setup, where the simulation is perfectly matched between the early non-equilibrium thermalization process and the ensuing hydro stage, one would expect that the overall dilepton spectra should be independent to the switching time. The reason for this is naturally that the pre-equilibrium stage flows smoothly into the hydrodynamical -- close-to-equilibrium -- stage. Nevertheless, in all state-of-the-art simulations  there will be regions of the space-time evolution, like cells in the boundary of the fluid, which do not perfectly satisfy the matching condition ($\tilde{\omega}\geq 1$). 
While we have checked that the average of the cells -- weighted by energy --  satisfies this condition, a large enough volume of fringe cells with $\tilde{\omega}_{\rm matching}<1$ may reduce the overlap. The reason for this is the parametric dependences of $\tilde{\omega}$, where increasing the shear viscosity, while fixing $\tau_{\rm hydro}$ and the temperature of the matched cell, $T_{\rm hydro}$, decreases the effective $\tilde{\omega}$, hence decreasing the overlap for smaller $\tau_{\hydro}$. In other words, the cell volume with $\tilde{\omega}_{\rm matching}<1$ becomes larger for earlier times.
We have quantified these effects in \cref{fig:hydro-dependenced},  where we plot the ratio of the total in-medium production spectrum (EKT+Hydro), for two different mathing times, $\tau_{\rm hydro}= [0.6,1.0]$. We have observed that for larger shear viscosities, the matching is less smooth. That being said,  such effects seem to be no larger than $20\%$ in magnitude for the systems studied in this work.

\bibliography{References}
\end{document}